\documentstyle[12pt]{article}
\begin{document}
\tolerance=5000
\def\be{\begin{equation}}
\def\ee{\end{equation}}
\def\bea{\begin{eqnarray}}
\def\eea{\end{eqnarray}}
\def\nn{\nonumber \\}
\def\cF{{\cal F}}
\def\det{{\rm det\,}}
\def\Tr{{\rm Tr\,}}
\def\e{{\rm e}}
\def\etal{{\it et al.}}
\def\erp2{{\rm e}^{2\rho}}
\def\erm2{{\rm e}^{-2\rho}}
\def\er4{{\rm e}^{4\rho}}
\def\etal{{\it et al.}}
\def\gsim{\ ^>\llap{$_\sim$}\ }

\ 

\vskip -2cm

\ \hfill
\begin{minipage}{3.5cm}
NDA-FP-58 \\
April 1999 \\
\end{minipage}

\vfill

\begin{center}
{\Large\bf 
Running gauge coupling and quark-antiquark potential 
from dilatonic gravity
}

\vfill

{\sc Shin'ichi NOJIRI}\footnote{\scriptsize 
e-mail: nojiri@cc.nda.ac.jp, snojiri@yukawa.kyoto-u.ac.jp} and
{\sc Sergei D. ODINTSOV$^{\spadesuit}$}\footnote{\scriptsize 
e-mail: odintsov@mail.tomsknet.ru, odintsov@itp.uni-leipzig.de}

\vfill

{\sl Department of Mathematics and Physics \\
National Defence Academy, 
Hashirimizu Yokosuka 239, JAPAN}

\ 

{\sl $\spadesuit$ 
Tomsk Pedagogical University, 634041 Tomsk, RUSSIA \\
and NTZ, Inst.Theor.Phys., University of Leipzig, 
Augustusplatz 10/11, 04109 Leipzig, Germany \\
}

\ 

\vfill

{\bf abstract}

\end{center}

The running gauge coupling and quark-antiquark potential 
in $d$-dimensions are calculated from the explicit solution of 
$d+1$-dimensional dilatonic gravity. This background interpolates 
between usual AdS in UV and flat space with singular dilaton in 
IR and it realizes two-boundaries AdS/CFT correspondence. The 
behaviour of running coupling and potential 
is consistent with results following from IIB supergravity.

\newpage

AdS/CFT correspondence \cite{W,Mal,GKP} predicts some properties 
of quantum gauge theory from higher-dimensional classical 
(super)gravity. In the more complicated versions of AdS/CFT 
correspondence there are two boundaries:
UV and IR boundaries \cite{BST,Be}.

In the recent paper \cite{NOA}, we presented two-boundaries AdS/CFT 
correspondence in dilatonic gravity (in the presence not only 
of metric tensor but also dilaton). The corresponding background 
\cite{NOA} interpolates between standard AdS at $y=\infty$ (UV) and
flat background with singular dilaton at $y=0$ (IR). The properties 
(in particular, conformal dimensions) for minimal
or dilaton coupled scalar around such background have been 
investigated.
Very similar background has been lately found in refs.\cite{KS,LT2}
(see also related refs.\cite{JPPZ}) 
for IIB supergravity. In these refs. 
the behaviour of running coupling in gauge theory and quark-antiquark 
potential \cite{Mal2} (with possible confinement due to dilatonic 
effects) has been studied. 
In particular, the modification of these quantities due to dilaton 
effects is explicitly found. Note that first example of 
running gauge coupling via bulk/boundary correspondence
has been presented for Type 0 strings in refs.\cite{KT,Mn}.
Background with singular dilaton under discussion 
gives another example of this beatiful phenomenon.

In the present note we calculate the running gauge coupling and 
quark-antiquark potential in $d$-dimensions working with our 
background \cite{NOA} 
realizing two-boundaries AdS/CFT correspondence in dilatonic gravity. 
We show that for some cases it may be similar to result of 
refs. \cite{KS,LT2}. 
Some clarifying remarks on interpretation of CFTs on two boundaries 
as IR/UV duality are also given.


We start with the following action in $d+1$ dimensions:
\be
\label{i}
S=-{1 \over 16\pi G}
\int d^{d+1}x \sqrt{-g}\left(R - \Lambda 
- \alpha g^{\mu\nu}\partial_\mu \phi \partial_\nu \phi \right)\ .
\ee
In the following, we assume $\lambda^2\equiv -\Lambda$ 
and $\alpha$ to be positive. 
In AdS/CFT correspondence dilatonic gravity with above action 
describes ${\cal N}=4$ super Yang-Mills theory superconformally 
interacting with ${\cal N}=4$ conformal supergravity. The corresponding 
conformal anomaly with non-trivial dilaton contribution has been found in 
ref.\cite{NOads} via AdS/CFT correspondence.

The equations of motion can be solved and the solution is given by 
\cite{NOA}
\bea
\label{v}
ds^2&=&\sum_{\mu,\nu=0}^d g_{\mu\nu}dx^\mu dx^\nu
=f(y)dy^2 + y \sum_{i,j=0}^{d-1}\eta_{ij}dx^i dx^j \\
\label{xii}
f&=&{d(d-1) \over 4y^2\left(
\lambda^2 + {\alpha c^2 \over y^d}\right)} \\
\label{xiii}
\phi&=&c\int dy \sqrt{{d(d-1) \over 
4y^{d+2}\left(\lambda^2 + {\alpha c^2 \over y^d}\right)}} \nn
&=&\phi_0+{1 \over 2}\sqrt{(d-1) \over d\alpha}\ln\left\{
{2\alpha c^2 \over \lambda^2 y^d}+1 \pm\sqrt{
\left({2\alpha c^2 \over \lambda^2 y^d}+1\right)^2 -1}\right\}\ .
\eea
Here $\eta_{ij}$ is the metric in the flat (Lorentzian) background 
and $c$ is a constant of the integration.
The boundary discussed in AdS/CFT correspondence lies at $y=\infty$.
 
In the string theory, the coupling on the boundary manifold, which 
would be the coupling in ${\cal N}=4$ $SU(N)$ super-Yang-Mills when 
$d=4$, is proportional to an exponential of the dilaton field $\phi$.
Although the relation of the present model with the string theory is 
not always clear (the model under discussion may be considered as 
low-energy string effective action), we assume the gauge 
coupling has the following form (by the analogy with 
refs.\cite{KS,LT2})
\be
\label{rg1}
g=g^*\e^{2\beta\sqrt{\alpha \over d(d-1)}\left(\phi-\phi_0\right)}\ .
\ee
Since the relation with the string theory is not clear, we put the 
constant coefficient $\beta$ in the exponent to be undetermined.
The factor $\sqrt{\alpha \over d(d-1)}$ and the shift of a constant 
$\phi_0$ are given for convenience and they may be always absorbed 
into the redefinition of $\beta$ and $g^*$, respectively.

The coupling $g$ is monotonically decreasing function and 
vanishes at $y=0$ when $\beta>0$ 
($\beta<0$) and $+$ ($-$) in the sign in (\ref{xiii}) are chosen.
On the other hand, the coupling is increasing one and diverges to 
$+\infty$ at $y=0$ for $\beta>0$ ($\beta<0$) and $-$ ($+$) sign. 
Therefore the $\pm$ sign would correspond to strong/weak coupling 
duality. When $y$ is large, which corresponds to asymptotically 
anti de Sitter space, $g$ behaves as
\be
\label{rg2}
g=g^*\left(1\pm {2\beta c\sqrt{\alpha} \over d\lambda y^{d \over 2}}
+ \left({2\beta \over d}-1\right){2\beta c^2 \alpha \over d 
\lambda^2 y^d} + \cdots \right)\ .
\ee
If we define a new coordinate $U$ by 
\be
\label{rg3}
y=U^2\ ,
\ee
$U$ expresses the scale on the (boundary) $d$ dimensional 
Minkowski space, which can be found from (\ref{v}).  
Following the correspondence between long-distances/high-energy 
in the AdS/CFT scheme, $U$ can be regarded as the energy scale of the 
boundary field theory. 
Then from (\ref{rg2}), we obtain the following 
renormalization group equation
\be
\label{rg4}
\beta(U)\equiv U{dg \over dU}= -d (g-g^*) 
- \left(d - {d^2 \over 2\beta} \right) {(g - g^*)^2 \over g^*}
+ \cdots\ .
\ee
The leading behaviour is identical with that in \cite{KS} 
for $d=4$ and the next to leading behavior of 
${\cal O}\left((g-g^*)^2\right)$ becomes also identical 
with theirs if we choose $\beta=4$. It maybe also 
noted that Type 0 strings on AdS background lead to
running gauge coupling as well \cite{Mn,KT}.

We now consider the static potential between ``quark'' and 
``anti-quark''. We evaluate the following Nambu-Goto action
\be
\label{rg5}
S={1 \over 2\pi}\int d\tau d\sigma \sqrt{\det\left(g^s_{\mu\nu}
\partial_\alpha x^\mu \partial_\beta x^\nu\right)}\ .
\ee
with the ``string'' metric $g^s_{\mu\nu}$, which 
could be given by multiplying a dilaton function $h(\phi)$ to 
the metric tensor in (\ref{v}). Especially we choose $h(\phi)$ 
by
\be
\label{rg6}
h(\phi)=\e^{2\gamma
\sqrt{\alpha \over d(d-1)}\left(\phi-\phi_0\right)}
= 1\pm {2\gamma c\sqrt{\alpha} \over d\lambda y^{d \over 2}}
+ \cdots \ .
\ee
Here $\gamma$ is an undetermined constant as $\beta$ in (\ref{rg1}). 
In order to treat general case, we assume $\gamma\neq \beta$ 
in general.
We consider the static 
configuration $x^0=\tau$, $x^1\equiv x=\sigma$, $x^2=x^3=\cdots
=x^{d-1}=0$ and $y=y(x)$.  Substituting the configuration into 
(\ref{rg5}), we find
\be
\label{rg7}
S={T \over 2\pi}\int dx h\left(\phi(y)\right) y \sqrt{
{f(y) \over y}\left(\partial_x y\right)^2 + 1}\ .
\ee
Here $T$ is the length of the region of the definition of $\tau$.
The orbit of $y$ can be obtained by minimizing the action $S$ 
or solving the Euler-Lagrange equation 
${\delta S \over \delta y}- \partial_x\left({\delta S 
\over \delta\left(\partial_x y\right)}\right)=0$. 
The Euler-Lagrange equation tells that 
\be
\label{rg8}
E_0={h\left(\phi(y)\right) y \over \sqrt{
{f(y) \over y}\left(\partial_x y\right)^2 + 1}}
\ee
is a constant. If we assume $y$ has a finite minimum $y_0$, where 
$\partial_x y|_{y=y_0}=0$, $E_0$ is given by
\be
\label{rg9b}
E_0=h\left(\phi(y_0)\right) y_0 \ .
\ee
 Introducing a parameter $t$, we parametrize $y$ by
\be
\label{rg9}
y=y_0\cosh t\ .
\ee
Then using (\ref{xii}), (\ref{rg6}), (\ref{rg8}) 
and (\ref{rg9}), we find
\bea
\label{rg10}
x&=&{y_0^{-{1 \over 2}} \over A}\int_{-\infty}^t dt 
\cosh^{-{3 \over 2}}t\left\{
1 \pm B\sinh^{-2}t\left(\cosh^2 t - \cosh^{2 - {d \over 2}}t\right)
+ {\cal O}(y_0^{-d})\right\} \nn
&& A\equiv{2\lambda \over \sqrt{d(d-1)}} \ ,\ \ \ 
B\equiv{2\gamma c\sqrt{\alpha} \over d\lambda} \ .
\eea
Taking $t\rightarrow +\infty$, we find the distance $L$ between 
"quark" and "anti-quark" is given by
\bea
\label{rg11}
L&=& {Cy_0^{-{1 \over 2}} \over A}\pm {BS_d y_0^{- {d+1 \over 2}} 
\over A} + {\cal O}(y_0^{- {2d+1 \over 2}}) \\
C&\equiv& \int_{-\infty}^\infty dt \cosh^{- {3 \over 2}}t
= {2^{3 \over 2} \Gamma\left({3 \over 4}\right)^2 
\over \sqrt{\pi}} \nn
S_d&\equiv&\int_{-\infty}^\infty dt \cosh^{{1 \over 2}}t 
\sinh^{-2}t \left(1 - \cosh^{- {d \over 2}} t \right) \ .\nonumber
\eea
Especially $S_4=C$. 
Eq.(\ref{rg11}) can be solved with respect to $y_0$ and we find
\be
\label{rg12}
y_0=\left({C \over AL}\right)^2\left\{ 1 \pm {BS_d \over C}
\left({AL \over C}\right)^d + {\cal O}\left(L^{2d}\right)\right\}\ .
\ee
 Using (\ref{rg8}), (\ref{rg9}) and (\ref{rg11}), we find the 
following expression for the action $S$
\bea
\label{rg13}
S&=&{T \over 2\pi}E(L) \nn
E(L)&=&\int_{-\infty}^\infty dt {dx \over dt}
{h\left(\phi(y(t))\right)^2 y(t)^2 \over 
h\left(\phi(y_0)\right)^2 y_0} \ .
\eea
Here $E(L)$  expresses the total energy of the 
``quark''-``anti-quark'' system.
The energy $E(L)$ in (\ref{rg13}), however, contains the divergence 
due to the self energies of the infinitely heavy ``quark'' and ``anti-quark''. 
The sum of their self energies can be estimated by considering the 
configuration $x^0=\tau$, $x^1=x^2=x^3=\cdots
=x^{d-1}=0$ and $y=y(\sigma)$ (note that $x_1$ vanishes here) 
and the minimum of $y$ is $y_0$. 
 Using the parametrization of (\ref{rg9}) and identifying $t$ 
with $\sigma$ ($t=\sigma$), we find the following expression of 
the sum of self energies:
\bea
\label{rg14}
E_{\rm self}=\int_{-\infty}^\infty dt\, h\left(\phi(y(t))\right)y(t)
\sqrt{f\left(y(t)\right)\left(\partial_t y(t)\right)^2 \over y}\ .
\eea
Then the finite potential between ``quark'' and 
``anti-quark'' is given by
\bea
\label{rg15}
E_{q\bar q}(L)&\equiv&E(L) - E_{\rm self} \nn
&=&\left({C \over AL}\right)\left\{D_0 \pm B 
\left(D_d+F_d+{S_d D_0 \over C}\right)
\left({AL \over C}\right)^d
+ {\cal O}(L^{2d})\right\} \nn
D_d&\equiv&2\int_0^\infty dt \cosh^{-{d+1 \over 2}}t\,\e^{-t} \\
&=&{2^{d -{5 \over 2}} \over (d-3)!!\sqrt{\pi}}
\Gamma\left({d-1 \over 4}\right)^2 - {4 \over d-1} \nn
F_d&\equiv&\int_{-\infty}^\infty dt \sinh^{-2}t\cosh^{1 \over 2}t
\left(1 - \cosh^{-{d \over 2}}t\right) \ .\nonumber
\eea
Especially since $F_4=D_4=C$, we obtain for $d=4$
\be
\label{rg15c}
E_{q\bar q}={\tilde L}^{-1}\left\{ - {2^{3 \over 2} \over 
\sqrt{\pi}}\Gamma\left({3 \over 4}\right)^2 + 4
\pm B\left({2^{1 \over 2} \over \sqrt{\pi}}\Gamma
\left({3 \over 4}\right)^2 + {8 \over 3}\right) {\tilde L}^4
+ {\cal O}({\tilde L}^8) \right\}\ .
\ee
Here
\be
\label{rg15d}
\tilde L \equiv {AL \over C}\ .
\ee
The behavior in (\ref{rg15c}) is essentially identical with the 
results in \cite{KS} but there are some differences. 
The coefficient in the leading term is different by an adding 
constant $4$, which  comes from the ambiguity when 
subtracting the self energy\footnote{Since $y_0$, the minimum of 
$y$, depends on the distance $L$ between ``quark'' and ``anti-quark'' 
as in (\ref{rg12}), the self-energy (\ref{rg14}) should depend on 
the distance $L$, which gives the ambiguity in the coefficient in 
the potential energy in (\ref{rg15}). }.
In the second term, there would be an ambiguity coming from the 
finite renormalization but by properly choosing the undetermined 
parameter $\gamma$ and the constant of the integration $c$ 
in (\ref{xii}), we can reproduce the result in (\ref{rg15c}). 

Let us try to understand better the background under consideration 
in relation with AdS/CFT correspondence. We have two boundaries 
at $y=0$ and $y=\infty$. 
Then the renormalization flow would connect two conformal 
field theories. 
To be specific, we only consider $d=2$ case in the following.
We also choose $-$ sign in $\pm$ in (\ref{xiii}) 
and $\phi_0=0$ since $\phi_0$ can be absorbed into the redefinition
of $G$.  We rescale the metric tensor by
\be
\label{rg16}
g_{\mu\nu}\rightarrow \tilde g_{\mu\nu}
=\e^{-2\sqrt{2\alpha} \phi}
g_{\mu\nu}\ .
\ee
 The redefined  metric behaves when $y\rightarrow \infty$ as
\be
\label{rg17}
d {\tilde s}^2\equiv \tilde g_{\mu\nu}dx^\mu dx^\nu 
= {1 \over 2\lambda^2 y^2} dy^2 + y \sum_{i,j=0}^{d-1}\eta_{ij}
dx^i dx^j \ . 
\ee
If we change the coordinate by $y=w^{-1}$, we obtain the standard 
metric on anti de Sitter space. On the other hand, the redefined 
metric behaves when $y\rightarrow 0$ as 
\be
\label{rg18}
d{\tilde s}^2={4\alpha c^2 \over \lambda^2}\left\{
{1 \over 2\alpha c^2y^2} dy^2 + y^{-1} \sum_{i,j=0}^{d-1}\eta_{ij}
dx^i dx^j \right\}\ . 
\ee
which is the metric of anti de Sitter space again.
In the redefinition of (\ref{rg16}), the action in (\ref{i}) is 
rewritten as follows:
\be
\label{rg19}
S=-{1 \over 16\pi G}
\int d^{d+1}x \sqrt{-g}\e^{-\sqrt{2\alpha}\,\phi}\left(R - 
\Lambda \e^{-2\sqrt{2\alpha}\,\phi}
+ 3\alpha g^{\mu\nu}\partial_\mu \phi \partial_\nu \phi \right)\ .
\ee
Therefore the effective Newton constant $\tilde G$ and the effective 
cosmological constant $\Lambda$ are given by
\be
\label{rg20}
\tilde G= \e^{\sqrt{2\alpha}\phi}G, \ \ 
\tilde \Lambda = \e^{-2\sqrt{2\alpha}\phi}\Lambda \ .
\ee
In the usual AdS/CFT correspondence, where the dilaton is 
constant, the central charge $c$ of the conformal field theory 
on the boundary is given by
\be
\label{rg21}
c={3l \over 2G}\ .
\ee
here the length scale $l$ is defined by 
$\Lambda = - {d(d-1) \over l^2}$. 
Similarly  defining $\tilde l$ by 
$\tilde\Lambda = - {d(d-1) \over {\tilde l}^2}$, we can define the
effective central charge by 
\be
\label{rg22}
\tilde c \equiv {3\tilde l \over 2\tilde G}=
{3l \over 2G}\ .
\ee
Note that $\tilde c$ is a constant everywhere, which would 
tell that the renormalization flow connects conformal field 
theories with same central charges. Therefore the conformal 
field theories on two boundaries would correspond to IR/UV 
duality (for a recent related discussion of RG flow 
see ref.\cite{FGPW}).

Finally let us note that one can easily generalize the present 
study for more general background containing other fields 
like gauge fields \cite{LT2}, antisymmetric tensor fields, etc.

\ 

\noindent
{\bf Acknoweledgements} The work by SDO has been partially supported 
by RFBR project N99-02-16617 and by Graduate College Quantum 
Field Theory at Leipzig University and by Saxonian 
Ministry of Science and Arts.

\end{document}